\documentclass[fleqn,usenatbib]{mnras}

\usepackage[T1]{fontenc}
\usepackage{ae,aecompl}

\usepackage{graphicx}      
\usepackage{amsmath}       
\usepackage{amssymb}       
\usepackage{braket}        
\usepackage{enumerate}     
\usepackage{dsfont}        
\usepackage{xcolor}        
\usepackage{algorithm}     
\usepackage{algpseudocode} 
\usepackage{ulem}          

\makeatletter
\def\@printed{
    \qquad\qquad\qquad Compiled using MNRAS \LaTeX\ style file v\@version}
\gdef\@journal{\hfill\@printed}
\def\bsp{}
\def\@oddfoot{}
 \def\@evenhead{}
 \def\@evenfoot{}
\def\ps@titlepage{\let\@mkboth\@gobbletwo
 \def\@oddhead{\footnotesize\@journal}
 \def\@oddfoot{\hfil}
 \def\@evenhead{\hfill}
 \def\@evenfoot{\hfil}
 \def\sectionmark##1{}
 \def\subsectionmark##1{}}
\def\abstract{\if@twocolumn
   \start@SFBbox\@abstract
 \else
   \@abstract
 \fi}
\def\endabstract{\if@twocolumn
   \endlist\finish@SFBbox
 \else
  \endlist
 \fi}
\def\@abstract{\list{}{%
    \listparindent\realparindent
    \itemindent\z@
    \labelwidth\z@ \labelsep\z@
    \leftmargin 1.5pc\rightmargin 1.5pc
    \parsep 0pt plus 2pt}\item[]%
    \reset@font\normalsize{\bf ABSTRACT}\\\reset@font\large
}
\def\@maketitle{\newpage
 \vspace*{7pt}
 {\raggedright \sloppy
  {\reset@font\huge \bf \@title \par}
  \vskip 23pt
  {\reset@font\LARGE
   \begin{tabular}[t]{@{}l@{}}\let\\=\author@nextline\@author
   \end{tabular}
   \par}
  \vskip 22pt
 }
 \par\noindent
 {\reset@font\small \par}
 \vskip 22pt
}
\makeatother

\title[]{Ridges in the Dark Energy Survey for cosmic trough identification}

\author[B. Moews et al.]{Ben Moews,$^{1}$\thanks{E-mail: \href{mailto:bmoews@roe.ac.uk}{bmoews@roe.ac.uk}}
Morgan A. Schmitz,$^{2}$
Andrew J. Lawler,$^{3}$
Joe Zuntz,$^{1}$
Alex I. Malz,$^{4}$
\newauthor
Rafael S. de Souza,$^{5}$
Ricardo Vilalta,$^{6}$
Alberto Krone-Martins,$^{7,8}$
Emille E. O. Ishida,$^{9}$\newauthor
for the COIN Collaboration 
\\
$^{1}$Institute for Astronomy, University of Edinburgh, Royal Observatory, Edinburgh EH9 3HJ, UK\\
$^{2}$Department of Astrophysical Sciences, Princeton University, 4 Ivy Ln., Princeton, NJ08544, USA\\
$^{3}$Department of Statistics, Baylor University, One Bear Place \#97140, Waco, TX 76798, USA\\
$^{4}$Ruhr-University Bochum, Astronomical Institute, German Centre for Cosmological Lensing,\\ \hspace{4pt}Universit{\"a}tsstr. 150, 44801 Bochum, Germany\\
$^{5}$Key Laboratory for Research in Galaxies and Cosmology, Shanghai Astronomical Observatory,\\ \hspace{4pt}Chinese Academy of Sciences, 80 Nandan Rd., Shanghai 200030, China\\
$^{6}$Department of Computer Science, University of Houston, 4800 Calhoun Rd., Houston TX, USA\\
$^{7}$Donald Bren School of Information and Computer Sciences, University of California, Irvine, CA 92697, USA\\
$^{8}$CENTRA/SIM, Faculdade de Ci\^encias, Universidade de Lisboa, Ed. C8, Campo Grande, 1749-016, Lisboa, Portugal\\
$^{9}$Universit{\'e} Clermont Auvergne, CNRS/IN2P3, LPC, F-63000 Clermont-Ferrand, France
}

\pubyear{2020}

\usepackage{newtxtext,newtxmath}

\begin{document}

\label{firstpage}
\pagerange{\pageref{firstpage}--\pageref{lastpage}}

\maketitle

\begin{abstract}
Cosmic voids and their corresponding redshift-projected mass densities, known as troughs, play an important role in our attempt to model the large-scale structure of the Universe. Understanding these structures enables us to compare the standard model with alternative cosmologies, constrain the dark energy equation of state, and distinguish between different gravitational theories. In this paper, we extend the subspace-constrained mean shift algorithm, a recently introduced method to estimate density ridges, and apply it to 2D weak lensing mass density maps from the Dark Energy Survey Y1 data release to identify curvilinear filamentary structures. We compare the obtained ridges with previous approaches to extract trough structure in the same data, and apply curvelets as an alternative wavelet-based method to constrain densities. We then invoke the Wasserstein distance between noisy and noiseless simulations to validate the denoising capabilities of our method. Our results demonstrate the viability of ridge estimation as a precursor for denoising weak lensing observables to recover the large-scale structure, paving the way for a more versatile and effective search for troughs. 
\end{abstract}


\raggedbottom

\section{Introduction}
\label{sec:introduction}

The cosmic web, a latticework structure of enormous proportions, represents one of the largest physical patterns in the Universe~\citep{Bond1996}. Its filamentary nature is a byproduct of the hierarchical growth of large-scale structure and gives rise to four main classes of substructures: Galaxy clusters, filaments, sheets, and the large regions of near-emptiness known as \textit{cosmic voids}~\citep{Zel1982}. 

These voids are characterised by underdensities in the dark matter distribution, presenting much simpler dynamics than their non-linear and high-density counterparts~\citep{Ham2016}. They are valuable cosmological probes, since they can encode relevant physical information and suffer from fewer sources of systematical error~\citep{Peebles2001, Lavaux2012}. Their population statistics may be predicted from a given cosmological theory~\citep[see][]{Fry1986, White1987, Li2011}, which can provide observational constraints to current models~\citep{Hoyle2004, Gruen2018}. They are especially useful for testing alternative cosmological models and probing screening mechanisms that are predicted to have a significantly reduced influence in such low-density regions, which we cover in more detail in \autoref{sec:discussion_applications}. However, in spite of their usefulness, the detection of voids is no trivial task. The challenge stems from the lack of a dominant definition thereof, and their detection remains a focal topic of interest in cosmology~\citep{Cai2015, Gruen2016, Sanchez2017, Nadatur2017, Adermann2018, Brouwer2018, 2019A&C....27...34X, Davies2020}.

Several established approaches to void detection employ Voronoi tessellation as described by~\citet{El-Ad1997} and~\citet{Gaite05}. Treated as particles, galaxies can, for example, be enclosed in distance-based Voronoi cells and grouped into larger zones, where a watershed simulation naturally leads to the identification of large basins (low-density regions) corresponding to voids~\citep{Neyrinck2008}. Markedly different approaches to void identification have also been proposed. For instance, \citet{2007A&A...474..315A} use computer vision techniques to classify void morphology, applying scale-independent morphology filters to identify primary cosmological structures such as walls, filaments, and voids. 

Alternatively, a different family of techniques exploits notions of data topology for void identification. \citet{2010ApJ...723..364A}, for example, apply topological segmentation, while~\citet{2019A&C....27...34X} apply notions of topological data analysis to find dimensional holes via persistent homology, in which zero, one, and two-dimensional holes are identified as clusters, loops of filaments, and voids, respectively. In the latter case, voids are considered statistically significant if their structure persists as long as data neighbourhoods increase in size. Classifications yielded by these techniques exhibit differences that impact the science case for which each method was developed~\citep{2018MNRAS.473.1195L}.

Despite the difficulties posed by automated void detection, the rich variety of current methods and techniques have lead to important advances in cosmology. As an example, the accurate location and modelling of voids can be exploited to derive clustering and abundance statistics such as void mass, and to constrain dark energy~\citep{Pisani2015}. The dynamics of matter flowing away from the centre of a void are instrumental to gain more insight on cosmological parameters, as described by~\citet{Dekel94}, while other applications include probing alternative dark matter models and tests of general relativity~\citep{Yang2015, Barreira2015}. 

Regarding the latter, \citet{Li2011} analyse void statistics in the context of scalar field theories in terms of their sizes, demonstrating through simulations that the the fifth force produced by scalar field coupling increases the fraction of large underdense regions and leads to sharper transitions between voids and filaments. They observe that this information can be used not only to establish constraints under $\Lambda$CDM, but also to distinguish between different coupled scalar field models. \citet{Cai2015} show that $f(R)$ gravity results in weaker gravitational lensing of voids due to their lower dark matter content, inducing differentiation from general relativity via the lensing tangential shear signal around voids. Similarly, void statistics such as abundances, ellipticities, radial density profiles, and radial velocity profiles have been used to study voids in simulations, leading to an increase in average void size and the elimination of small voids, as well as emptier voids~\citep{Zivick2015}. The sharper transitions may be detectable in the projected 2D ridges that this paper investigates.

Another difficulty inherent to void detection is the need for accurate redshift measurements. Reliable redshifts demand larger, and ideally complete, galaxy spectroscopic surveys, which tend to cover only small areas of the sky. Photometric galaxy surveys, such as the Panoramic Survey Telescope and Rapid Response System ~\citep[Pan-STARRS;][]{Chambers2016}, the Dark Energy Survey~\citep[DES;][]{flaugher}, and the upcoming Vera C. Rubin Observatory Legacy Survey of Space and Time~\citep[LSST;][]{Ivezic08}, on the other hand, can provide observations covering larger areas.

Beyond these studies of cosmic voids, or the more common study of the galaxy clusters that occupy the nodes of the cosmic web, the characterisation of the filaments that connect them can also be of great interest. The filament between clusters Abell 0399 and 0401, for example, has been shown to host quiescent galaxies and hot gas, and to emit in radio~\citep{bonjean2018, govoni2019}. Automated detection of filaments in a large volume has recently been carried out by~\citet{malavasi2020}, using galaxy samples from the Sloan Digital Sky Survey~\citep[SDSS;][]{York2000}, while~\citet{galarraga2020} apply a similar approach to hydrodynamical simulations to derive the expected statistical properties of filaments.

In these studies, the distributions of galaxies from spectroscopic surveys (or simulations thereof) are used to detect filaments. Alternatively, weak gravitational lensing can be used as the observable to extract filaments~\citep{mead2010,maturi2013}. While a few individual detections have been reported between specific clusters, as described by~\citet{dietrich2012} and~\citet{jauzac2012}, the very low signal-to-noise ratio of the filament signal typically requires the stacking of large numbers of pairs of clusters to make detection from weak lensing possible~\citep[see, for example,][and references therein]{xia2020}. The very low amplitude of the lensing signal of filaments poses a significant challenge to detection efforts from wide photometric surveys and using lensing observables. Even if this were not the case, any such attempt to characterise many filaments over a large 3D volume would suffer from the same dependency on redshift accuracy as that discussed in the case of studies focussed on voids.

A way to avoid relying on accurate redshift measurements, while still readily taking advantage of photometric surveys, is to consider the 2D-projected counterparts to the elements that make up large-scale structure instead~\citep[though see also][for a void finder built to work on photometric surveys, using redshift slices instead of a full 2D projection]{Sanchez2017}. In the case of voids, such projections are known as \textit{troughs}, which represent the most underdense regions in the plane of the sky. Since troughs comprise regions of lower density across the line of sight in the projected space only, it is sufficient to use photometric measurements, obviating spectroscopic redshifts~\citep{Clampitt2013,Gruen2016}.

In related research, \citet{Brouwer2018} make use of the photometric Kilo-Degree Survey~\citep[KiDS;][]{deJong2017} and the spectroscopic Galaxy And Mass Assembly Survey~\citep[GAMA;][]{Driver2011} to identify troughs from foreground galaxies; their simulations ultimately forecast that upcoming surveys such as LSST and Euclid will enable us to constrain the redshift evolution of cosmic troughs. This finding is further supported by research showing 2D-projected underdensities, meaning troughs, to be a powerful probe to distinguish between $\Lambda$CDM and modified gravity models in future lensing surveys, potentially even more so than 3D cosmic voids~\citep{Higuchi2016, Barreira2017, Cautun2018}.

In this work, we propose an algorithm to detect 2D density ridges as a way to denoise cosmic structure in mass density maps. For this purpose, we extend the subspace-constrained mean shift (SCMS) algorithm introduced by~\citet{Ozertem2011} to fit our application case, and apply our method to the DES Y1 data release~\citep{flaugher}. The general methodology works by defining a denoised and sparse representation of the filamentary structure, and as a consequence, the locations of regions of emptiness emerge naturally. Considering the aforementioned limitations of spectroscopic surveys, trough finders can be applied as an alternative to void finders to recover and study underdense regions from weak lensing studies, giving particular relevance to the present application to DES Y1.

While previous work demonstrates how to identify filaments as density ridges using the SCMS algorithm~\citep[see][further discussed in \autoref{sec:methodology}]{Chen2015a,Chen2015b,Chen2016}, our method extends past implementations in several ways. We incorporate the haversine distance, a more suitable approach for spherical surfaces, and also customise the mesh size for ridge estimation and optimisation of the bandwidth. Another important difference is that these studies apply the algorithm to galaxy data, either from simulations or SDSS, or directly to dark matter particles from N-body simulations. In this work, we apply the approach to 2D weak lensing mass maps instead; the ridges we recover are thus extracted from the projected matter distribution and not directly comparable to the filaments of the cosmic web. Finally, we compare our ridges to a search based on curvelets, an extension of the wavelet transform more suitable for the study of curvy and filamentary structures~\citep{candes2006}. 

This paper is organised as follows. 
\autoref{sec:methodology} explains the SCMS algorithm and implemented extensions; it includes our kernel density estimation technique, experimental data, and simulations. \autoref{sec:results_and_discussion} describes our experimental results and compares them to a curvelet-based denoising technique. \autoref{sec:discussion} comments on our approach and future directions, with an emphasis on the advantages and drawbacks of the proposed methodology. Finally, \autoref{sec:conclusion} provides a summary and conclusions. 

\section{Methodology and data}
\label{sec:methodology}

This section provides background information on the SCMS algorithm in \autoref{sec:subspace-constrained_mean_shift} and describes past applications, as well as extensions from both this and prior work, in \autoref{sec:extensions_of_the_core_algorithm}. Lastly, we introduce the Dark Energy Survey and its mass maps, together with our approach to sample generation and the creation of noisy and noiseless simulations for verification purposes in \autoref{sec:dark_energy_survey_density_maps}.

\subsection{Subspace-constrained mean shift}
\label{sec:subspace-constrained_mean_shift}

Introduced by~\citet{Ozertem2011}, the SCMS algorithm is a recent addition to statistical methods dealing with the estimation of density ridges. Starting with a mesh of points placed in equidistant steps across the parameter space, the algorithm seeks to establish local principal curves in iterative steps. This can be visualised as a cloud of points shifting closer towards the nearest underlying structure at each iteration, akin to the process in which mass in our universe converges toward better-defined cosmic filaments over time. The latter can be observed in N-body simulations such as the Millennium Simulation by~\citet{Springel2005a} and its Millennium-II successor by~\citet{Boylan2009}, as well as the Bolshoi simulation by~\citet{Klypin2011} and the MultiDark simulation~\citep{Riebe2013}.

In more formal terms, a ridge is a maximiser of the local density in the normal direction as given by the Hessian matrix. Let $\nabla p(x)$ be the gradient of a probability density function $p$ on a space of dimension $d$, $H(x)$ its Hessian matrix of second derivatives, and $v$ the eigenvectors of $H(x)$ corresponding to a descending sorting of eigenvalues $\lambda$. We can diagonalise $H(x)$ according to
\begin{eqnarray}
H(x) = U(x) \lambda(x) U(x)^{\top}.
\end{eqnarray} 
We then take the $d - 1$ eigenvectors (columns of $U(x)$) corresponding to the $d - 1$ smallest eigenvalues $\lambda$ as $v'$, thereby omitting the column for the largest eigenvalue, which corresponds to the direction parallel to the ridge. Taking these eigenvectors and their linear projection operator,
\begin{eqnarray}
L(x) \propto L(H(x)) = v' v'^{\top},
\end{eqnarray}
we can then project the gradient of $p$ onto the eigenvectors as
\begin{eqnarray}
G(x) = L(x) \nabla p(x).
\end{eqnarray}
Following the deeper investigation of nonparametric ridge estimation methods of \citet{Genovese2014}, a ridge $R$ can thus be expressed as
\begin{eqnarray}
R = \{ x : ||G(x)|| = 0, \lambda_{d + 1} (x) < 0 \},
\end{eqnarray}
with $x$ as the locations in which $G$ is zero everywhere and the omitted largest eigenvalue is positive\footnote{Without this second condition, we also locate valleys.}. 

Pseudocode describing the SCMS algorithm as it appears in~\citet{Moews2019a} can be found in Algorithm~\ref{alg:scms}, including thresholding, with the notation of $\theta_{\ast, 1}$ and $\theta_{\ast, 2}$ defining all rows of the first and second column of an $N$-by-2 matrix, respectively. Line 4 shows a kernel density estimation with a radial basis function (RBF) kernel, $\mathcal{K} (x) = (1 / \sqrt{2 \pi}) \exp(-0.5 x^2)$, meaning
\begin{eqnarray}
\label{eq:kde_rbf}
\mathrm{KDE}_{\mathrm{RBF}} (x, \beta) = \frac{1}{\mathrm{dim}(\theta) (2 \pi \beta^2)^{\frac{d}{2}}} \sum\limits_{i = 1}^{\mathrm{dim}(\theta)} e^{\left( \frac{|| x - \theta_i ||^2}{2 \beta^2} \right)}.
\end{eqnarray}

\begin{algorithm}
\caption{SCMS with thresholding as in~\citet{Moews2019a}}
\begin{algorithmic}[1]
\State \textbf{Input:} Coordinates $\theta$, bandwidth $\beta$, threshold $\tau$, iterations $N$
\State \textbf{Output:} Density ridge point coordinates $\psi$
\Procedure {SCMS}{$\theta$, $\beta$, $\tau$, $N$}
\State $\kappa(x) \longleftarrow \mathrm{KDE}_{\mathrm{RBF}} (\theta, \beta)$, using Eqn.~\ref{eq:kde_rbf}
\State $x_{\mathrm{min}} \longleftarrow (\min(\theta_{\ast, 1}), \max(\theta_{\ast, 1}))$
\State $y_{\mathrm{min}} \longleftarrow (\min(\theta_{\ast, 2}), \max(\theta_{\ast, 2}))$
\State $\mathcal{\psi} \longleftarrow \mathcal{\psi} \sim U(x_{\mathrm{min}}, y_{\mathrm{min}})_{\mathrm{dim}(\theta)}$
\State $\psi \longleftarrow \forall y \in \psi : \kappa(y) > \tau$
\For {$n \longleftarrow 1, 2, \dots, N$}
\For {$i \longleftarrow 1, 2, \dots, \mathrm{dim}(\psi)$}
\For {$j \longleftarrow 1, 2, \dots, \mathrm{dim}(\theta)$}
\State $a_j = \frac{\psi_i - \theta_j}{\beta^2}$
\State $b_j = \mathcal{K} \left( \frac{\psi_i - \theta_j}{\beta} \right)$
\EndFor
\State $H(x) = \frac{1}{\mathrm{dim}(\theta)} \sum_{j = 1}^{\mathrm{dim}(\theta)} b_j \left( a_j a_j^{\top} - \frac{1}{\beta^2} \mathbb{I} \right)$
\State $v, \lambda \longleftarrow v, \lambda \mathrm{ \ from \ diagonalisation \ eig} (H(x))$
\State $v' \longleftarrow \mathrm{entries \ in \ } v \mathrm{ \ corresp. \ to \ sort}_{\mathrm{asc}} (\lambda)_{1, 2, \dots, d - 1}$
\State $\psi_i \longleftarrow v' v'^{\top} \frac{\sum_{j = 1}^{\mathrm{dim}(\psi)} b_j \theta_j}{\sum_{j = 1}^{\mathrm{dim}(\psi)} b_j}$
\EndFor
\EndFor
\State \textbf{return} $\psi$
\EndProcedure
\end{algorithmic}
\label{alg:scms}
\end{algorithm}

\subsection{Previous applications and extensions}
\label{sec:extensions_of_the_core_algorithm}

In the few years since its introduction, the SCMS algorithm has proven valuable in a variety of fields, from the analysis of 3D neuron structures in tissue images by~\citet{Bas2011} to the identification of road networks in satellite images by combining the SCMS algorithm with the geodesic method and tensor voting~\citep{Miao2014}. 

\subsubsection{Applications to astronomy and thresholding}
\label{sec:astronomy_and_thresholding}

The first applications of the SCMS algorithm in astronomy are also the ones that are most closely related to our work. For the purpose of investigating galaxy evolution, and after initially using SDSS data as a test dataset for a methodological paper by~\citet{Chen2014}, \citet{Chen2015a} employ the algorithm to constrain matter distributions at different redshifts by applying \textit{thresholding}, the practice of discarding low-density areas of the initial grid of unconverged ridge points according to a kernel density estimate of the data, precluding their identification as filaments. More formally, the threshold $\tau$ is determined by computing the root mean square of the differences between the average density estimate and the grid points' density estimates,
\begin{eqnarray}
\label{eq:threshold}
\tau = \sqrt{\frac{\sum (\mathbf{\phi - \bar{\phi}})^2}{G}},
\end{eqnarray}
for an element-wise subtraction of an array of grid point density estimates $\mathbf{\phi}$ with average $\bar{\phi}$ over $G$ grid points. In a related paper, they also explore the algorithm's suitability to show that dark matter is traced by baryonic matter across large-scale structure~\citep{Chen2015b}. Additionally, \citet{Chen2016} provide a filament catalogue for SDSS, and show the influence of nearest-filament distances on galaxy properties like size, colour, and stellar mass. 

Another application, this time in cosmology, investigates non-Gaussianities of the matter density field to provide lensing effects based on filaments using the SCMS algorithm~\citep{He2017}. \citet{Hendel2018} gain a better understanding of galactic mergers through left-over collision disruptions by using the algorithm to classify stellar debris and identify morphological substructures therein.

The most recent use of the algorithm pertains to the field of quantitative criminology, where multiple extensions are introduced~\citep{Moews2019a}. The study investigates the optimisation of police patrols via density ridges, using publicly available data from the City of Chicago over multiple years to assess the validity and stability of predictive ridges. Apart from thresholding as described above, the study makes use of the haversine formula as described by~\citet{Inman1835}, which calculates the great-circle (or orthodromic) distance as a way to prevent distorted measures that would result from, for example, using the Euclidean distance. As a more specialised case of the law of haversines, it computes the distance between two points along the surface of a sphere. 
The formula makes use of the haversine function for a given angle $\alpha$, 
\begin{eqnarray}
\label{eq:haversine_function}
\mathrm{hav} (\alpha) = \frac{1 - \cos(\alpha)}{2} = \sin^2 \left( \frac{\alpha}{2} \right),
\end{eqnarray}
and can be used to calculate the relative haversine distance between two such points, $\theta_1$ and $\theta_2$, as
\begin{eqnarray}
\begin{aligned}
\label{eq:haversine_distance}
\delta_{\mathrm{hav}} (\theta_1, \theta_2) = \mathrm{hav} (&\theta_{2, 1} - \theta_{1, 1} \\
&+ \cos \theta_{1, 1} \cos \theta_{2, 1} \mathrm{hav} (\theta_{2, 2} - \theta_{1, 2})).
\end{aligned}
\end{eqnarray}
The resulting ridge estimation tool by~\citet{Moews2019a} is publicly available as a python package called \textsc{DREDGE}\footnote{\url{https://pypi.org/project/dredge}}, with a corresponding open-source repository\footnote{\url{https://github.com/moews/dredge}}. 

In this paper, we adapt and extend \textsc{DREDGE} in order to apply it to DES Year 1 mass density maps. In addition to the existing thresholding and integrated use of the haversine formula, we parallelise time-consuming parts of the code, enable the manual setting of the mesh size for the ridge estimation, and implement a mathematically grounded optimisation of the required bandwidth. The latter is due to~\citet{Moews2019a} using a simple automatic bandwidth calculation suited to the initial application in criminology. Some of the features in the original implementation of \textsc{DREDGE} are redundant for this paper as well, most notably the ability to set a top-percentage threshold to extract only ridges falling within the highest-density areas, as researchers and practitioners in criminal justice are often interested in focussing on `hot spots'~\citep{Braga2005}.

For the purpose of parallelising \textsc{DREDGE}, we make use of the embarrassing parallelism inherent in the updating function of the ridge points at each iteration. 
As this update is not reliant on other ridge points during the respective iteration, multiprocessing offers an easily accessible option to speed up the algorithm's runtime. 

\subsubsection{Kernel density estimation}
\label{sec:kde}

Kernel density estimation is a well-developed non-parametric data smoothing technique that has seen wide use in cosmological research applications. \citet{park2007}, for example, use an adaptive smoothing bandwidth with a spline kernel, and~\citet{mateus2007} apply a $k$-nearest neighbours density estimator to estimate the local number density of galaxies in an SDSS sample.

Similar methods have been employed in other surveys, for example by~\citet{scoville2007}, who identify large-scale structure and estimate dimensions, number of galaxies, and mass using an adaptive smoothing technique over a sample in the Cosmic Evolution Survey~\citep[COSMOS;][]{Scoville2007_2}. Relatedly, \citet{jang2006} uses a multivariate kernel density estimator with a cross-validated smoothing parameter to estimate galaxy cluster density over a sample of the Edinburgh-Durham Cluster Catalogue (EDCC).

The SCMS algorithm's bandwidth $\beta$, plays a crucial role in determining the bias-variance relationship of the resulting distribution. A larger bandwidth results in a smoother distribution with less variance and more bias, whereas a smaller bandwidth results in a less smooth distribution with more variance and less bias. Finding an optimal bandwidth for the kernel estimator in the context of DES mass maps is crucial to ensure that dense regions will not be oversmoothed and that higher-density areas of the projected large-scale structure will not be blurred into troughs. Conversely, optimising the bandwidth allows us to preserve the properties of low-density and extended structures. 

In this application, we use a likelihood cross-validation approach to find the optimal bandwidth parameter. This method provides a density estimate that is close to the actual density in terms of the Kullback-Leibler divergence~\citep[KLD;][a measure of relative entropy]{Kullback1951}. Cross-validation is performed using a maximum likelihood estimation of the leave-one-out kernel estimator of $f_{-i}$, which is given by
\begin{eqnarray} \label{eq:looke}
f_{-i}(X_{i})=\frac{1}{(n-1)h} \sum_{j=1,j\neq i}K_{h}(X_{i},X_{j}),
\end{eqnarray}
where $h$ is the bandwidth parameter and $K_{h}$ represents the generalised product kernel estimator. We use the generalised product kernel estimator of~\citet{Li2006} on the latitude and longitude coordinates of the DES Y1 data as
\begin{eqnarray} \label{eq:gpke}
K_{h}(X_{i},X_{j}) = \prod_{s=1}^{q}h_{s}^{-1}k\left(\frac{X_{is}-X_{js}}{h_{s}}\right),
\end{eqnarray}
where $q$ is the dimension of $X_i$, $X_{is}$ is the $s^{\mathrm{th}}$ component of $X_i$ ($s = 1, \dots , q$), $h_s$ is the smoothing parameter for the given component of $X_i$, and $k(\cdot)$ is a univariate kernel function. This nonparametric kernel estimator does not assume any functional form of the data, only that it satisfies regularity conditions such as smoothness and differentiability. In our case, the $\beta$ value we use in SCMS is the best $h$ value obtained by cross-validation, while the variables $X_i,X_j$ correspond to the sky positions $\theta$.

\subsection{Data and simulations}
\label{sec:dark_energy_survey_density_maps}

\subsubsection{DES mass maps}
\label{sec:des_mass_maps}

The Dark Energy Survey (DES) is a six-year photometric survey project to image 5000 square degrees of the sky in grizY filters using the DECam camera on the Blanco telescope, Cerro Tololo, Chile~\citep{flaugher}. The primary purpose of the survey is to generate a dataset for cosmology, and in particular one suitable for weak gravitational lensing measurements. DES observations completed in January 2019, and data analysis for the project is ongoing.

Weak lensing measurements use galaxies as a backlight to determine the projected gravitational fields along the paths of their light rays. Gravity bends the light paths, resulting in a shearing of galaxy images that can be measured from the mean ellipticity of a large-enough sample of objects. One application of weak lensing is to generate mass maps. Assuming the general relativity relationship between gravitational convergence $\kappa$ and mass, we can use ellipticity catalogues to map the projected, weighted overdensity in a given pixel of the survey. The weighting depends on the redshift distribution of the observed galaxies and the redshift-distance relationship that, in turn, depends on the underlying cosmology.

The DES Year 1 (Y1) data release\footnote{\url{https://des.ncsa.illinois.edu/releases/y1a1}}, as described by~\citet{drlica-wagner}, includes 2D-projected mass maps~\citep[see][]{chang} estimated from cosmic shear measurements from the survey's first year~\citep{zuntz}. The creation of these maps is based on the redshifts estimated in \citet{hoyle} and the shear catalogues made using the {\sc Metacalibration} method~\citep[see][]{HuffMandelbaum2017,SheldonHuff2017} which inverts shears to convergences by applying the spherical Kaiser-Squires method to the galaxy shear catalogues~\citep{kaiser-squires, schneider}.

Various mass maps were made using different selections of source galaxies, thus having different redshift weight functions. For this initial project, we use only the maps made with the widest range of galaxies, from $z = 0.2$ to $z = 1.3$. Here, we use the E-mode maps and their corresponding masks\footnote{\url{http://desdr-server.ncsa.illinois.edu/despublic/y1a1_files/mass_maps}, files \texttt{y1a1\_spt\_mcal\_0.2\_1.3\_kE.fits} and \texttt{y1a1\_spt\_mcal\_0.2\_1.3\_mask.fits}}.

\subsubsection{Sample generation from DES maps}
\label{sec:betasamp}

Although the mass maps are already in the form of a field on which Hessians and gradients may be calculated, in practice, the DES Y1 mask, which has a large number of small excised regions at the locations of bright stars, makes calculating derivatives a very noisy process, even with aggressive masking. It is considerably simpler to instead generate samples from the map and apply the SCMS algorithm described above. In order to generate these samples, we compute a mean value $\mu_i$ per pixel,
\begin{eqnarray}
\mu_i = \text{max}(1 + \omega \kappa_i, 0),
\end{eqnarray}
where $\kappa_i$ is the projected overdensity in pixel $i$, and $\omega$ is a parameter that we can tune as desired. If $\omega$ is too low, the ridges will not be detectable in the map, and if it is too high, the lower ridges will disappear because the highest density peaks dominate. We find that a value of $\omega = 50$ works well for DES data to suppress the $\sim 2\%$ map fluctuations to within $o(1)$ point density variation. We then generate a number $n_i$ of samples per pixel, using a Poisson distribution $n_i \sim \textrm{Poi}(\mu_i)$, and place these samples uniformly within the pixel.

Note that no interpolation or reconstruction is performed in masked regions; masked pixels are omitted, and no samples are generated within them. This method is simple to apply and causes no numerical difficulties, but it does mean that ridges at the edge of the mask regions may not be detected. This occurs only in regions where interpolation would also fail, as the SCMS algorithm, like reconstructive methods, is able to identify structure across missing pixels. Methods using these ridge catalogs should account for this, for example in simulations.

\subsubsection{Flask simulations}
\label{sec:verification_through_simulations}

To provide a testbed for our method before employing real data, we also build a suite of simulated maps using the {\sc Flask}\footnote{\url{https://github.com/hsxavier/flask}} software~\citep[see][]{flask, xavier}, which generates tomographic log-normal random fields that approximate large-scale structure distributions. We use the Planck best-fit $\Lambda$CDM cosmological parameters in the code, $\sigma_e = 0.25$, and the same redshift distribution as estimated for the DES source galaxies~\citep[see][]{hoyle}, normalised to the correct overall density. We then generate true $\kappa$ maps, which we treat as idealised noiseless simulations, and galaxy ellipticity catalogues, which we use with a spherical Kaiser-Squires map-making method to generate a noisy $\kappa$ map. Lastly, we apply the DES masks to both noisy and noiseless simulated maps. 

While this approach works well for the experiments performed in this work, it is insufficient for pseudo-3D extensions of our method discussed in Section~\ref{sec:discussion_applications}, as {\sc Flask} cannot generate features like clusters and filaments.

\section{Experimental results}
\label{sec:results_and_discussion}

In this section, we discuss and implement a distance-based statistical test to verify our results, and we use noisy and noiseless simulated dark matter density maps in \autoref{sec:simulation-based_verification} to test the degree of robustness to noise. 
We explore the general and specific properties of our method's tracing of the large-scale structure through a quantifiable comparison to the curvelet transforms in \autoref{sec:comparison_with_curvelet_transforms}. Lastly, we present the extracted ridges, together with a comparison to previous research on trough identification from DES Y1 data, in \autoref{sec:ridges_in_the_dark_energy_survey}.

\subsection{Statistical functionality verification}
\label{sec:simulation-based_verification}

As is common among studies dealing with cosmic voids, validating our approach is not an easy task, even when we can apply it to simulations. Indeed, while simulations allow us to apply our method to noiseless data, they do not provide us with forward-modelled `ground-truth' ridges or troughs, since these need to be estimated and defined, even in the absence of noise. A viable analysis is the comparison of ridges and troughs recovered when running the proposed approach in the diverse settings of a noiseless simulation and one that contains realistic levels of noise, as described in \autoref{sec:verification_through_simulations}. In other words, we can test for robustness to observational noise. 

Such a test, however, requires the choice of a similarity criterion, or distance metric, between sets of ridge points, in this case those extracted from noisy and noiseless maps. If such a metric could be defined on the ridges themselves, it could be applied directly to the ridges recovered from the proposed approach. This would allow us to directly probe the approach's sensitivity to observational noise, without requiring any specification of how to relate the ridges to a scientifically meaningful quantity like cosmological parameters. While such a test is not sufficient to guarantee that the ridges contain the information required to compute such quantities, it can confirm that the ridges are not dominated by observational noise, a necessary if not sufficient condition for any science case, provided that an appropriate metric is chosen.

Optimal transport theory~\citep[see][]{villani2008}, and specifically the Wasserstein distance, provides us with a natural, principled means of computing such distances. Per the Monge-Kantorovich interpretation of optimal transport, the Wasserstein distance can be understood as the minimal possible cost incurred to move a certain amount of mass from one distribution to another. This is precisely what a set of ridges and troughs represent; a certain distribution of mass, projected in two dimensions across (a part of) the sky.

Consider two distributions of mass, $p_1, p_2 \in \mathbb{R}^N$, sampled on some discrete space of dimension $N$. Let $C\in\mathbb{R}^{N\times N}$ be the cost matrix whose entries $C_{ij}$ contain the cost of moving mass from position $i$ to position $j$. We define the Wasserstein distance as
\begin{eqnarray}\label{eq:wass}
W(p_1, p_2) = \mathrm{argmin}_{T\in \Pi(p_1,p_2)} <T,C>.
\end{eqnarray}
Here, $\Pi(p_1,p_2)$ is the set of transport plans between $p_1$ and $p_2$. For any such matrix $T$, each of its entries $T_{ij}$ contains the amount of mass of $p_1$ that is transported from position $i$ within $p_1$ (denoted $p_{1,i}$) to position $j$ within $p_2$. By construction, for any row $i$, we then have
\begin{eqnarray}
\sum_j T_{ij} = p_{1,i}.
\end{eqnarray}
Similarly, summing over the columns of $T$ yields the entries of $p_2$. We can then simply express the set of acceptable transport plans as
\begin{eqnarray}
\Pi(p_1,p_2) = \left\{T\in\mathbb{R}^{N\times N}, \forall i,j, \sum_j T_{ij} = p_{1,i}, \sum_i T_{ij} = p_{2,j}\right\}.
\end{eqnarray}
In order to find the optimal $T$, the solution to Eqn. \eqref{eq:wass}, we use the entropic regularisation scheme proposed by~\citet{cuturi2013}, which allows for the Wasserstein distance between discrete measures to be computed by using an iterative scheme~\citep[see the textbook by][for a recent overview of computational schemes for the practical computation of optimal transport quantities]{peyre2019}.

\begin{figure}
\centering
\includegraphics[width=\linewidth]{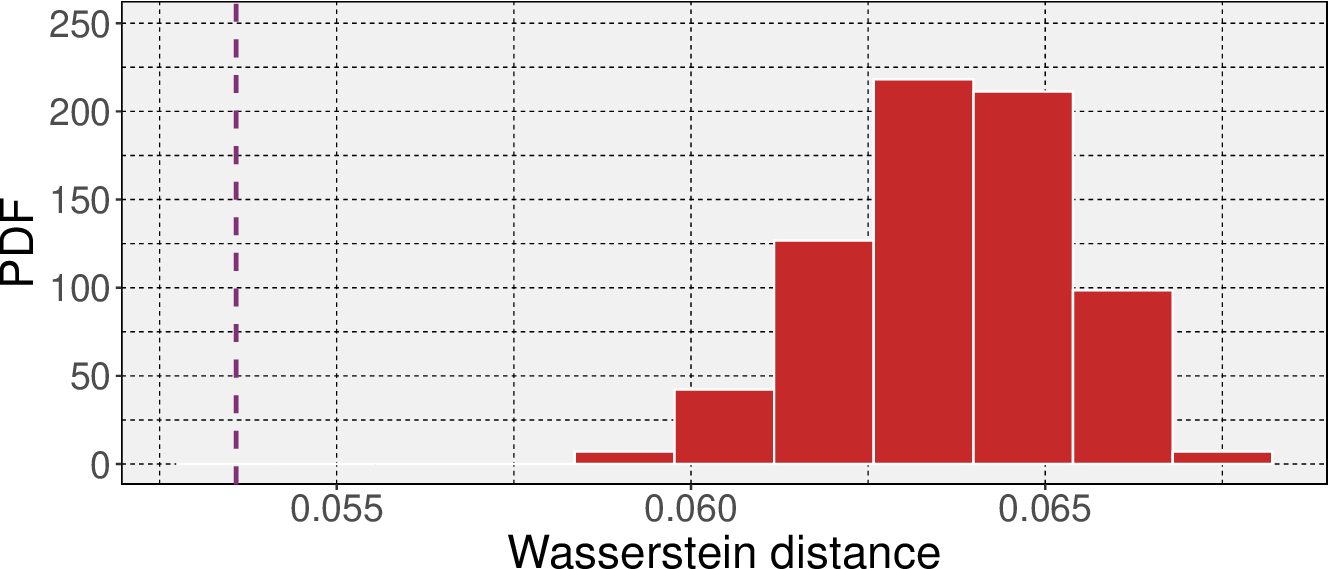}
\caption{Wasserstein distance between the ridges obtained on the noisy simulation and either its noiseless counterpart, as a vertical dashed line in purple, or a set of 101 random distributions of mass in red.}
\label{fig:wassdist}
\end{figure}

The raw output of SCMS is a set of shifted `mesh points', each with a 2D position vector, that make up the ridges. While we could compute the Wasserstein distance directly between two sets of mesh points, the objects of interest in our case are the ridges they comprise, and the troughs thus delimited, rather than the points themselves. For this reason, we convert each set of ridge points into a binary 2D image with each pixel's value set to zero if no mesh point fell within it, and one otherwise. We then compute the Wasserstein distance between the two resulting images. Here, $N$ is equal to the total number of pixels, in our case $10952$, and the cost matrix $C$ is the Euclidean distance between each pair of pixel positions. The use of the latter distance metric is motivated by comparing binarised 2D images, as opposed to within-image calculations on the curved sky as part of the SCMS algorithm.

In order to provide a basis for comparison, we generate random maps by projecting the DES mask onto the image plane and uniformly sampling the same number of non-zero pixels as those present in the images that contain our ridges. We then compute the Wasserstein distance between those ridges and the ones obtained from the noisy simulation. The distribution of these distances is represented in \autoref{fig:wassdist}, along with the distance computed between the two sets of ridge points based on noisy and noiseless maps. As can be seen, our ridge-finding method shows robustness to realistic amounts of noise. This test allows for a general confirmation that the ridges we obtain contain physical information about the distribution of matter, obviating any specification of the problem under study, that is, a precise definition of troughs or voids. As mentioned above, further testing tailored to the application of interest is recommended. These tests could likely use the same settings and simulations, where we compare the output in both noiseless and noisy cases, for example the cosmological constraints derived from the troughs delimited by those two sets of ridges.

\subsection{DES ridges and curvelet comparison}
\label{sec:comparison_with_curvelet_transforms}

\begin{figure*}
\includegraphics[width=0.995\linewidth]{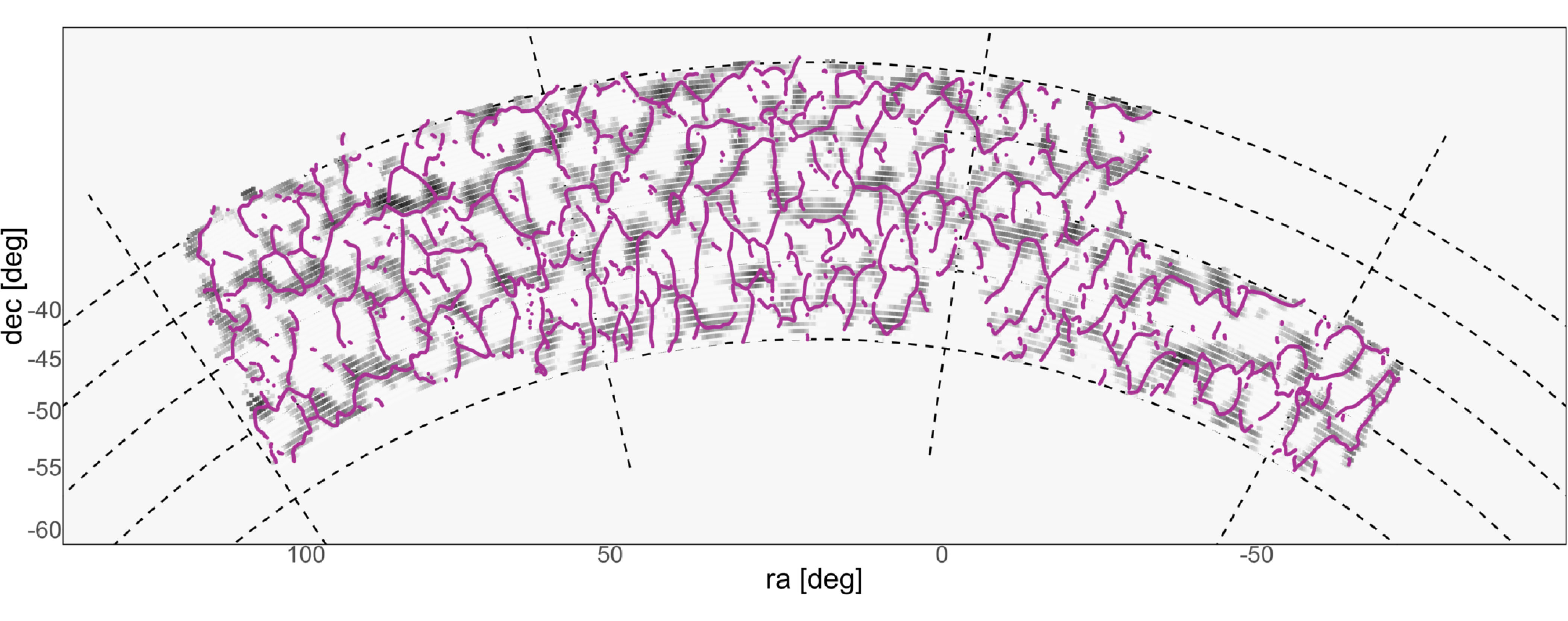}
\caption{Comparison of density ridges and a curvelet reconstruction. Ridges in purple are superimposed on structural constraints obtained via curvelet denoising in shades of grey, with higher densities shifting from lighter to darker. Both results are based on DES Y1 weak lensing mass density maps.}
\label{fig:curvelets}
\end{figure*}

Sparse signal processing~\citep[see][]{starck2015} provides solutions to many signal retrieval problems including, but not exclusive to, image denoising. The approach relies on finding a representation space in which the signal can be sparsely represented, a typical example being a sinusoidal signal, which can be fully expressed with very few non-zero coefficients in Fourier space. While natural signals are rarely sinusoidal, wavelets are commonly used as a sparse basis of representation~\citep{mallat1999}. They can, however, perform poorly when the features to be recovered are rectilinear or elongated, as is the case for estimating ridges. Because of this shortcoming, analogous transforms have been designed specifically for these cases, namely ridgelets and curvelets~\citep{candes1999, candes2006}. 

These transforms have led to a wide range of applications in astrophysics and cosmology~\citep{starck2003}. \citet{starck2004} use wavelets, ridgelets and curvelets to detect and characterise, on simulations, various sources of CMB anisotropies, which include imprints of inflation, the Sunyaev-Zel'dovich effect, and cosmic strings. The latter have also been studied in a CMB framework by~\citet{vafaeisard2017} and~\citet{hergt2017}, using the curvelet transform, while~\citet{laliberte2018} use ridgelets to that end within N-body simulations. \citet{gallagher2011} apply both to solar astrophysics, and~\citet{jiang2019} use curvelets for radio transient detection.

In our case, the application of curvelets allows for a straightforward and entirely independent means of recovering the ridges. We perform a simple denoising, that is, a thresholding of the input mass in curvelet space. Our SCMS-recovered ridges are obtained from the samples described in \autoref{sec:betasamp}, which themselves rely on the choice of the $\omega$ parameter. We use these samples to generate a two-dimensional, discrete image by counting their number in each pixel bin. We then apply curvelet denoising to the resulting image, using the freely available \texttt{Sparse2D} package\footnote{\url{https://github.com/CosmoStat/Sparse2D}}. The thresholds are selected using the False Discovery Rate approach described by~\citet{benjamini1995}, and the coarse scale is discarded. The resulting denoised map, converted back to sky coordinates, is shown in \autoref{fig:curvelets}. Upon visual inspection, we observe good agreement between the uncovered structured overdensities and the ridges obtained by SCMS, overplotted in purple. 

\begin{figure*}
\includegraphics[width=0.995\linewidth]{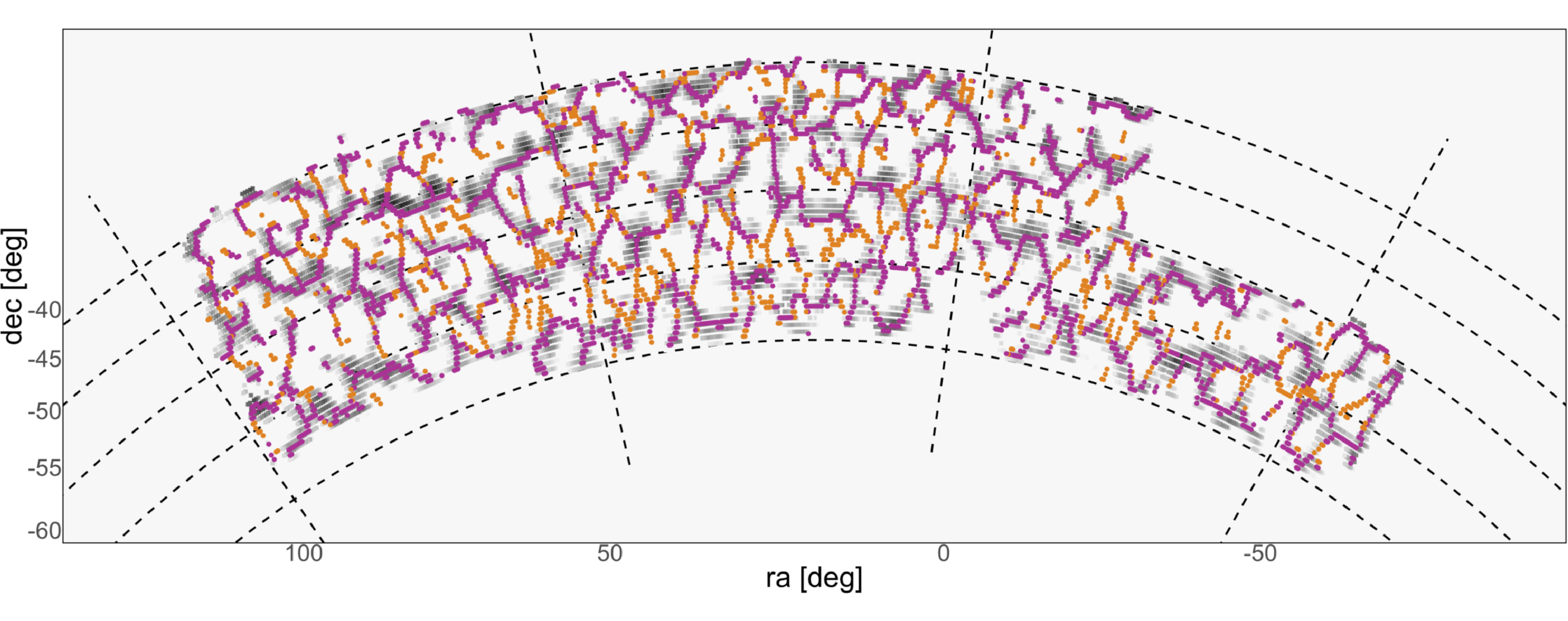}
\caption{Similar to \autoref{fig:curvelets}, but with ridges shown in purple where they match the curvelet reconstruction, and orange otherwise.}
\label{fig:redVsGreen}
\end{figure*}

To get a more quantitative view of the differences in the structures recovered by both approaches, we project the ridges yielded by the SCMS algorithm into the same pixel grid as that used by the curvelet denoising step. For every pixel that then contains part of one of the ridges, we check the corresponding value in the curvelet reconstruction. We consider all pixels where that value is $0$ to be a mismatch. \autoref{fig:redVsGreen} shows all such mismatches in orange, while parts of the ridges that match with the curvelet reconstruction are shown in purple.
About $31\%$ of the SCMS-derived ridges mismatch the curvelet reconstruction. 

As can be seen from \autoref{fig:redVsGreen}, the mismatches mostly correspond to areas where structure is present in two nearby areas of the sky, and where the curvelet reconstruction keeps those two areas disjoint, while SCMS-derived ridges link them together. Although very different in their heuristics, both approaches ultimately perform some form of denoising of the input mass maps. Those differences between the two denoising approaches could indicate that one of the two fails to properly separate noise from signal; the areas of disagreement could either be due to the curvelets considering signal to be noise, or to SCMS algorithm turning noise into ridges. 

\begin{figure} 
\includegraphics[width=0.995\linewidth]{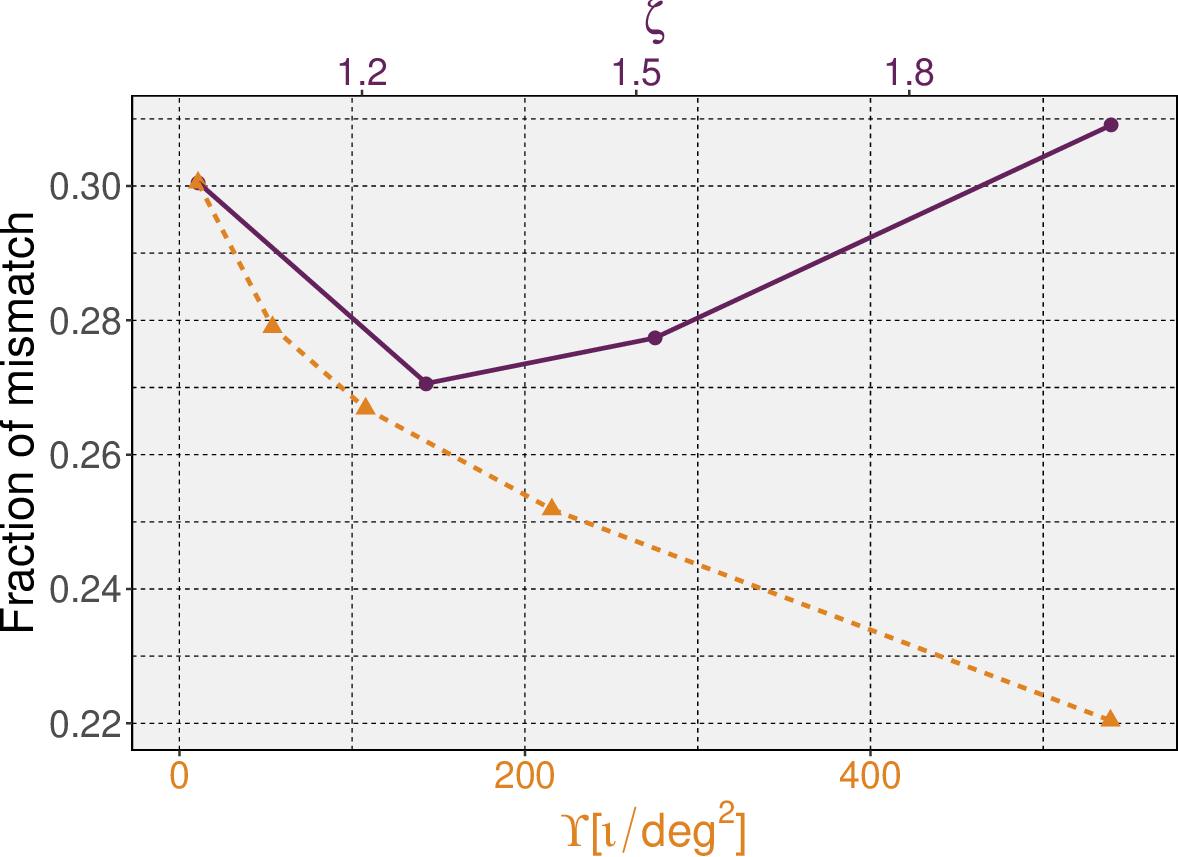}
\caption{Fraction of `mismatch' between SCMS-derived ridges and the curvelet reconstruction, when varying the $\tau$ and $\beta$ hyperparameters of the SCMS algorithm. The bottom axis shows the threshold on the ridges ($\Upsilon$) in units of meshpoints ($\iota$) per square degree (deg$^2$). The top axis shows the multiplication factor ($\zeta$) of the bandwidth. The further to the right, the stronger the implicit denoising performed by the algorithm. The absence of a clear decreasing trend shows that the differences between both methods, as illustrated by \autoref{fig:redVsGreen}, are not due to hyperparameter choices.}
\label{fig:hyperparams}
\end{figure}

In order to root out the cause of these discrepancies, we reprocess the DES data while tuning the parameters of each method. In the case of SCMS, we impose a stronger denoising by multiplying the bandwidth, $\beta$, obtained as described in \autoref{sec:kde}, by a factor $\zeta \in \{1.25, 1.50, 2.00\}$, or considering a higher threshold $\Upsilon > \tau$ (see \autoref{sec:astronomy_and_thresholding}). The percentage of mismatching ridges for both cases is shown in \autoref{fig:hyperparams}. 

Artificially increasing the strength of the denoising performed by the SCMS algorithm does not lead to ridges that match the curvelet reconstruction more closely. In fact, the percentage of mismatch between the two approaches is not even monotonic with respect to the bandwidth. In all cases, we still observe ridges that tend to be more connected in the SCMS case. Similarly, we try imposing weaker denoising in the curvelet approach, by using increasingly lower values of $k$ in $k\hat\sigma$ thresholds instead of the False Detection Rate approach, where $\hat\sigma$ is estimated from the data using the Median Absolute Deviation estimator. Once again, this yields no clear increase in the match between the two outputs.

This shows that the support of each method, in their respective (hyper)parameter spaces, are disjoint. In other words, the differences between curvelet reconstruction and SCMS-derived ridges seen in \autoref{fig:redVsGreen} are due to intrinsic differences between the approaches, as opposed to a failure of either at the denoising task. Implications of those differences for potential applications are discussed further in \autoref{sec:discussion_applications}.

\subsection{Ridges in the Dark Energy Survey}
\label{sec:ridges_in_the_dark_energy_survey}

\begin{figure*}
\includegraphics[width=0.995\linewidth]{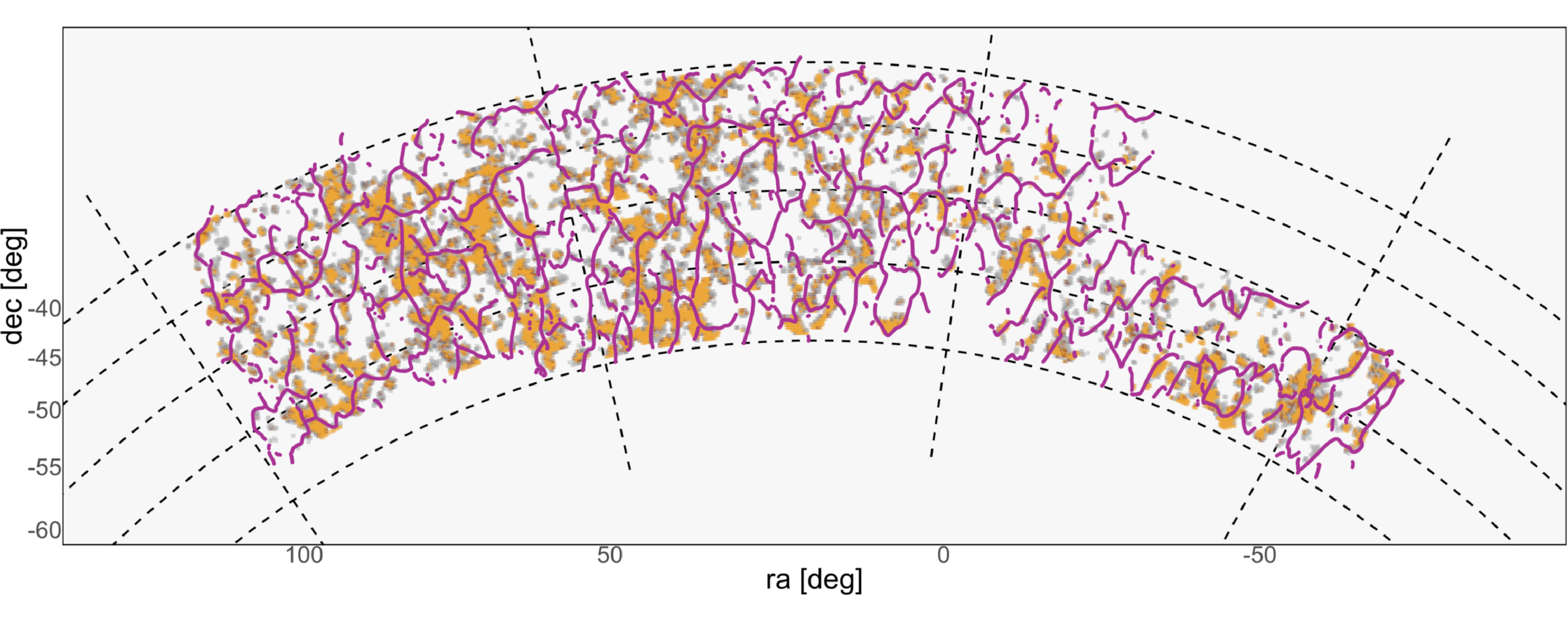}
\caption{Comparison of density ridges and previous results from~\citet{Gruen2018}. Ridges from this work are shown in purple, and are superimposed on mass density probabilities that were obtained by measuring counts-in-cells along lines of sight of the foreground luminous red galaxies \textsc{redMaGiC} sample.}
\label{fig:gruen_comparison}
\end{figure*}

After the above comparison with curvelet transforms, we perform a second comparison of our results with independent methodology, as void and trough detection remain a current focus of interest within the cosmology community, as described in~\autoref{sec:introduction}. Specifically, \citet{Gruen2018} derive cosmological constraints via density split statistics, counting tracers in cells to split lines of sight and measuring counts-in-cells and gravitational shear in regions of varying density. Using \textsc{redMaGiC} luminous red galaxies at $0.2 < z < 0.45$ to trace the foreground matter density field, they count the number of galaxies that fall within circular top-hat apertures within radii $\theta_T = [10', 20', 30', 60']$.

Each line of sight is then assigned to a density quintile based on those counts, the data for which is publicly available\footnote{\url{https://des.ncsa.illinois.edu/releases/y1a1/density}, file \texttt{trough\_maps.tar.gz}}, with the highest quintile being of interest to us in terms of high-density ridge analogues. Due to the DES Y1 data being more inhomogeneous in depth than the SDSS DR8 data used for the \textsc{redMaGiC} catalogue~\citep[see][]{Aihara2011}, tracer galaxies are removed and the area is defined as fully masked if the sample from the catalogue is not complete to $z = 0.45$. We compare our findings to~\citet{Gruen2018}, which uses the same DES Y1 data we make use of in this work in combination with SDSS DR8 data, without sharing the same underlying mass density map, as in the curvelet comparison. \autoref{fig:gruen_comparison} shows the same DES Y1-extracted ridges as in \autoref{fig:curvelets}, underlaid with the highest-quintile density measurements from~\citet{Gruen2018}.

While these quintile-based comparisons are more `spotty' due to the masking based on depth completeness described above, and qualitatively different when compared to curvilinear structures extracted by the SCMS algorithm, the lines generally trace the same high-density regions shown in the figure. The shown percentiles also use foreground galaxies as tracers of the matter field, as opposed to the mass density maps based on weak lensing that are used as the input for our modified version of \texttt{DREDGE}. In contrast to the previous comparison with curvelets as an alternative method in \autoref{sec:comparison_with_curvelet_transforms}, the goal of this experiment is to perform a comparison across both methods and underlying data to reach a consensus in terms of high-density regions on the sky.

\section{Discussion}
\label{sec:discussion}

This section provides an in-depth discussion of our findings, their implications, and future direction. In Section~\ref{sec:discussion_overview}, we recapitulate the implemented extensions and performed experiments, discuss the results of the latter, provide an overview of advantages and disadvantages, and compare the SCMS algorithm to ridge-like structures extracted by other means. In Section~\ref{sec:discussion_applications}, we describe planned follow-up research for developing and testing cosmological probes and models, and to validate cosmological analysis pipelines.

\subsection{Overview}
\label{sec:discussion_overview}

Ridges derived from our methodology have several properties that make them interesting for lensing trough studies, compared to simply using local minima. First, since the method generates a point cloud as a starting point, rather than working with the map directly, it does not require convergence map smoothing. This means that smaller-scale troughs can potentially be probed. Secondly, since the trough points maximise the distance from local ridge structures, they probe inter-cluster and inter-filament regions directly, rather than by proxy. Finally, masks from bright stars on small scales (below the typical ridge segment length) do not significantly affect the operation of the algorithm.

We incorporate several extensions from previous research into our implementation of the SCMS algorithm, both for accuracy and performance reasons, and describe these in \autoref{sec:extensions_of_the_core_algorithm}. Thresholding the initial mesh of points based on density estimates, as introduced by~\citet{Chen2015a}, ensures that the curvilinear structure is constrained to higher-density areas, which solves potential issues associated with identifying ridges in sparsely populated regions in the data. We also make use of the haversine formula, as implemented by~\citet{Moews2019a} for geospatial analysis, to prevent the distortion of measurements on the curved sky.

We further extend the algorithm by exploiting the potential for embarrassing parallelism inherent in the updating function of ridge points at each iteration. This allows users to reduce the runtime significantly by using a multiprocessing setup, thus removing a major obstacle when applying the SCMS algorithm to large datasets. The algorithm itself is reliant on the choice of a bandwidth, which plays a crucial role in determining the bias-variance relationship of the distribution, with larger bandwidths resulting in smoother distributions with less variance and more bias, and with smaller bandwidths resulting in a less-smooth distribution with more variance and less bias. For our current application, this means that large-scale structure in dark matter density maps could be blurred into cosmic troughs through bandwidths that are too large, while recovered ridges could show spurious fine-grained structure through bandwidths that are too small. We solve this by introducing a likelihood cross-validation to the SCMS algorithm to automatically find the optimal bandwidth in a data-driven way, providing a density estimate close to the actual density in terms of the Kullback-Leibler divergence.

The experiments performed in this work are designed to test for both noise robustness and the correct tracing of high-density regions. In terms of noise, we generate noisy and noiseless simulations that correspond to the DES Y1 mass density maps in \autoref{sec:simulation-based_verification}, and calculate the Wasserstein distance between binarised maps of the resulting curvilinear structures. In addition, we calculate the same metric for randomly generated maps with uniform sampling to place the results into context, showing robustness to realistic levels of noise. While this test provides some measure of the robustness of our method to noise in real data, as discussed in \autoref{sec:simulation-based_verification}, additional testing is recommended to avoid introducing biases specific to any application. However, as our goal is to propose a general approach to recover ridges, our tests were chosen to be as general as possible.

The overall agreement we find between our ridges and the structure recovered by curvelet denoising in \autoref{sec:comparison_with_curvelet_transforms}, despite the independence and vast difference between the two approaches, is a good indication that our ridges successfully capture information contained in the matter distribution. The differences between the two methods lie mostly in the curvelet reconstruction leading to more disconnected patches, while our ridges tend to connect these areas. Our experiments show that this is the result of intrinsic differences between both approaches rather than a poor choice of hyperparameters; while clearly different, neither of the two resulting estimates are `wrong'. This illustrates an important point: For a given matter distribution, there is no such thing as a set of true ridges. Choosing a proper definition is a non-trivial task in itself, which is precisely one of the motivations for the present work.  In this paper, where we aim to present both methods in as general a framework as possible, we will simply point out the difference in the structures identified by each, highlighting once again that neither is more correct than the other. Tests tailored to a specific application, however, could be used to determine which of the two is more appropriate or useful in that context. 

Another key difference between the two approaches is that the SCMS algorithm does indeed produce ridges, that is, curvilinear structures, whereas the output of the curvelet denoising is still a full two-dimensional map. The curvelet reconstruction also still contains information about the amplitude of the signal at each position, while our ridges are binary, meaning that every position either contains a ridge or not. If such a truly curvilinear format is required for a given application, and the patchier nature of the curvelet-recovered `ridges' is better suited for the respective task at hand, it would be straightforward to combine both approaches (see \autoref{fig:redVsGreen}).

To further our analysis, we compare ridges extracted from DES Y1 mass density maps, which are based on weak lensing, to trough mass probabilities by \citet{Gruen2018} in \autoref{sec:ridges_in_the_dark_energy_survey}, using the highest quintile corresponding to large-scale structure. Since this test uses \textsc{redMaGiC} luminous red galaxies at $0.2 < z < 0.45$ rather than weak lensing to trace the foreground matter density field and is also limited to the depth coverage of the combined DES Y1 and SDSS DR8 data, it serves as a comparison across both methods and sources. Given this combination in~\citet{Gruen2018}, a stronger masking is present, leading to a more `spotty' nature of the visible structure when compared to the curvelet representation in \autoref{sec:comparison_with_curvelet_transforms}. Despite these differences, we see both ridges and quintile probabilities falling into the same areas, with DES Y1-extracted ridges tracing this complementary dataset.

Our cross-validated bandwidth optimisation approach to kernel density estimation provides a data-driven and, via the Kullback-Leibler divergence, mathematically motivated way to perform ridge estimation. As with any approach that is data-driven, sanity checks should be performed to ensure physically plausible ridges. These derived ridges are, therefore, quite dependent on both the data quality of the DES Y1 survey and the shape assumptions in the density estimation, as well as on the distribution space being approximately symmetric and unimodal. In our approach, care should be taken when considering samples derived from heavy-tailed distributions, as the bandwidth optimisation can be prone to overestimation~\citep{silverman1986density}. To encourage both verification and further analysis of this denoising approach to dark matter density maps, we release a catalogue\footnote{The data table is available at CDS via anonymous ftp to \texttt{cdsarc.u-strasbg.fr} (130.79.128.5) or via \href{http://cdsarc.u-strasbg.fr/viz-bin/qcat?J/MNRAS}{http://cdsarc.u-strasbg.fr/viz-bin/qcat?J/MNRAS}.} of the ridges extracted from DES Y1 data.

\subsection{Applications}
\label{sec:discussion_applications}

In terms of future directions and follow-ups, we intend to extend our method to the three-dimensional case to identify full voids instead of trough projections. One way to reach this goal is to use tomographic 2D information, effectively identifying voids within a layered pseudo-3D structure. This can be performed using tomographic reconstruction techniques~\citep[see][]{Gabor09}, which have been highly enhanced using deep neural networks~\citep{Wang18}.

Another interesting avenue we plan to pursue is the inclusion of lensing information along identified ridges to further bolster the viability of this approach, as well as the matching of a cosmic web reconstructed from DES data to Planck LSS~\citep{Bouchet16}. This follow-up study will also continue the work of~\citet{Gruen2016} on weak lensing shear around troughs to further probe the connection between convergence fields and matter density, and extend research on optimal trough finders for weak lensing maps by~\citet{Davies2020}. For the latter, we propose an iterative expanding-circle approach to trough identification, qualitatively similar to the spherical void finder by~\citep{Padilla2005}, as the preparatory step of finding troughs in our catalogues. Ridges such as the ones presented in this work are especially useful for this, as the lack of a smoothing requirement of the convergence map allows for the potential identification of smaller-scale structure, and thus troughs. By inducing a sparse, hence denoised, representation of ridges structures, our approach is also particularly useful for the topological analysis of ridge structures via, for example, persistent diagrams and Betti numbers~\citep{2019A&C....27...34X}.

An interesting avenue of follow-up research is the differentiation between the standard model and alternative cosmologies. Specifically, cosmologies such as scalar-tensor theories and massive gravity employ screening mechanisms for compatibility, decoupling a proposed fifth force from matter living in high-density regions like galactic interiors and making the fifth force a function of environment~\citep{Desmond2019}. These screening mechanisms would have the least effect in highly underdense regions, making voids an ideal test bed for investigations. The approach for tomographic pseudo-3D voids outlined above could, thus, be easily adapted to test alternative cosmologies. \citet{Davies2019} find that troughs, through abundances and weak lensing profiles, can also act as sensitive probes of gravity, opening up a direct extension of the presented work to tests of cosmological models.

Insofar as the ridges we locate here are projections of filaments (which remains to be shown) and other structures intermediate between voids and clusters, lensing along and around them also offers a potential probe of screening mechanisms: We expect screening to switch off at some point as we move away from the ridge, and this may be detectable. A stacking approach akin to that presented in \cite{xia2020}, but around ridges, should prove informative on these poorly understood structures, along with other tracers of ridge topology and density. Any prediction code used to obtain likelihoods needs to account for the strong noise biases due to identifying ridges via gravitational lensing, and then measuring lensing around those same ridges. The most likely method for theory predictions of such quantities is interpolating between simulations, as is done in most peak and trough analyses. Consequently, this means that simulations must have very accurate noise properties, as opposed to being homogeneous.

In addition to the testing potential of voids for cosmological models outlined in \autoref{sec:introduction}, \citet{Barreira2015} investigate cubic Galileon and nonlocal gravity theories as modified gravity pathways that change the lensing potential. In terms of screening, the former theory does not screen such modifications inside of voids, featuring lensing effects roughly twice as strong as predicted by general relativity, offering an avenue for lensing-based cosmological tests. For the Vainshtein mechanism, a specific screening mechanism native to a variety of gravity models, \citet{Falck2018} also note that dark matter in large-scale structure is not screened by the latter, and demonstrate that voids are completely unscreened and can be used to show deviations from the $\Lambda$CDM model through tangential velocity and velocity dispersion of stacked voids. Regarding void stacking, \citet{Cautun2016} also offer a non-spherical stacking approach based on a boundary profile, which could further enhance such tests. These previous results offer convincing evidence that a number of void and trough statistics can be used effectively to test for alternative cosmological models, especially those making use of screening mechanisms.

 This planned future application would yield a three-dimensional void catalogue, the statistical properties of which could constitute powerful cosmological probes. To start, the higher-order statistics of the spatial distribution of voids could be compared to results from~\citet{Hamaus14} and~\citet{Pycke16}. Additional insights could also be gained from the spatial correlation of voids as a function of their sizes and shapes. These void statistics also open up possibilities as a sensitive probe of the dark energy equation of state, as demonstrated by~\citet{Pisani2015} for the number of observed voids and~\citet{Demchenko2016} for their profiles in the context of the spherical evolution model, possibly to be combined with complementary constraints on the dark energy equation of state~\citep{Moews2019b}. Conversely, \citet{Bos2012} find that morphological properties of voids are too strongly affected by sparsity and spatial bias in their given sample to differentiate cosmologies at a statistically significant level.

Just as the statistical characterisation of the triaxiality of galaxy clusters has proven fruitful for cosmological analysis~\citep[see][]{2017MNRAS.466.3103S, Melchior_2017, 2018ApJ...860..126C}, so too might the triaxiality distribution of voids. One could further imagine a hierarchical inference procedure that uses the void size and shape distributions as a prior to iteratively update ambiguous photometric redshift probability density functions, just as those same probability density functions are used to hierarchically infer the void size and shape distributions from the tomographic projections used to derive the triaxial void catalogue. A supplemental application of these distributions could serve as a validation test for the sophisticated simulated catalogues being developed for future LSS surveys~\citep{Korytov19}.

In short, this method brings within reach a number of promising avenues for testing cosmological models, developing novel cosmological probes, and validating cosmological analysis pipelines. In this paper, we demonstrate the utility of curvilinear structures for the denoising of large-scale structure based on weak lensing, extending the available methodology for this purpose. We plan to explore these applications in future papers, but also welcome potential users of the ridge catalogue to experiment with the dataset we provide alongside the presented results.

\section{Conclusion}
\label{sec:conclusion}

This work presents a new ridge estimation approach based on an extension of the subspace-constrained mean shift algorithm, a filamentary search method, and releases the corresponding results as a catalogue of curvilinear structure. As an application case of current relevance, we apply the method to dark matter mass density maps from the DES Y1 data release to extract high-density ridges between cosmic troughs. Our results demonstrate the viability of ridge estimation as a precursory step for denoising cosmic filaments, leading to a versatile and effective identification of cosmic troughs.

We extend the SCMS algorithm by including the haversine distance, a customisation of the mesh size for ridge estimation, automatic optimisation of the bandwidth used in the process, and the parallelisation of the updating function for mesh points to scale down the algorithm's runtime. We also include the thresholding extension of the SCMS algorithm from a previous application to astronomical data. In order to test the robustness of our method, we recover the ridges from simulations under different noise levels, and use the Wasserstein distance as a comparison metric.

We further compare our extracted ridges, which are based on DES Y1 weak lensing data, with curvelet denoising of the same data, and with high-density quintiles derived from both DES Y1 and SDSS DR8 foreground galaxies limited by inhomogeneous depth coverage. This allows us to compare results across both methods and data sources, leading to highly reasonable agreement. Lastly, we discuss the utility of our approach in the context of further investigations, with a focus on ridge lensing, testing alternative cosmological models through troughs and pseudo-3D voids, and screening mechanisms.

\section*{Acknowledgements} 
This work was developed during the $\rm 6^{th}$ COIN Residence Program\footnote{\url{https://cosmostatistics-initiative.org/residence-programs/crp6}} (CRP\#6) held in Chamonix, France in August 2019. This project is financially supported by CNRS as part of its MOMENTUM programme over the 2018--2020 period. The Cosmostatistics Initiative\footnote{\url{https://cosmostatistics-initiative.org}} (COIN) is a non-profit organization whose aim is to nourish the synergy between astrophysics, cosmology, statistics, and machine learning communities. AIM acknowledges support from the Max Planck Society and the Alexander von Humboldt Foundation in the framework of the Max Planck-Humboldt Research Award endowed by the Federal Ministry of Education and Research. AKM acknowledges the support from the Portuguese Funda\c c\~ao para a Ci\^encia e a Tecnologia (FCT) through grants SFRH/BPD/74697/2010, PTDC/FIS-AST/31546/2017 and from the Portuguese Strategic Programme UID/FIS/00099/2013 for CENTRA.

\section*{Data availability}
The data underlying this article will be made available in the VizieR database of astronomical catalogues at the \href{http://cds.u-strasbg.fr/}{Centre de Donn{\'e}es astronomiques de Strasbourg (CDS)} website, accessible via \texttt{http://cdsarc.u-strasbg.fr/viz-bin/cat/J/MNRAS/X/Y}, with X and Y as the final volume and page number of this publication, respectively.

\bibliographystyle{mnras}
\bibliography{ref}

\bsp
\label{lastpage}

\end{document}